\documentclass{aa}
\usepackage{graphics,epsfig}
\newcommand{\chem}[2]{$\rm{}^{#1}\kern-0.8pt#2$}
\newcommand{\chim}[2]{\rm{}^{#1}\kern-0.8pt#2}
\newcommand{\reac}[6]{$\rm\,{}^{#1}\kern-0.8pt{#2}\,({#3}\,,{#4})\,
           {}^{#5}\kern-0.8pt{#6}\,$}

\begin{document}

\title{Actinides: How well do we know their stellar production?}

\author{S. Goriely and M. Arnould} 
\institute{Institut d'Astronomie et d'Astrophysique, Universit\'e Libre de Bruxelles,
B-1050 Brussels, Belgium}

\date{Received date; accepted date}
  
\abstract{
The reliable evaluation of the r-process production of the actinides and careful estimates of
the uncertainties affecting these predictions are key ingredients especially in
nucleo-cosmochronology studies based on the analysis of very metal-poor stars or on the
composition of meteorites. This type of information is also required in order to make
the best possible use of future high precision data on the actinide composition of galactic
cosmic rays, of the local interstellar medium, or of meteoritic grains of presumed circumstellar
origin. This paper provides the practitioners in these
various fields with the most detailed and careful analysis of the r-process actinide production
available to-date. This study is based on a version of the multi-event canonical model of the
r-process which discards the largely used waiting point approximation. It considers also different
combinations of models for the calculation of nuclear masses, $\beta$-decay  and
fission rates. Two variants of the model used to
predict nuclear reaction rates are adopted. In addition, the influence of the level
of Pb and Bi production by the r-process on the estimated actinide production is 
evaluated by relying on the solar abundances of these two elements. In total, thirty-two
different cases are presented, and are considered to give a fair picture of the level of
reliability of the predictions of the actinide production, at least in
the framework of a simple r-process model. This simplicity 
is imposed by our inability to identify the proper astrophysical sites for the r-process.
As a guide to the practitioners,  constraints on the actinide yield predictions and
associated uncertainties are suggested on grounds of the measured abundances of
r-nuclides, including Th and U, in the star CS 31082-001, and under the critical and
questionable assumption of the `universality' of the r-process. We also define alternative
constraints based on the nucleo-cosmochronological results derived from the present actinide content
of meteorites. Both sets of constraints suffer from serious
problems. The first set does not hold in the likely situation of the non-universality of the
r-process. The definition of the second set is made difficult by the necessity of using intricate
galactic evolution models in order to interpret the meteoritic data. Our calculations of the actinide
production combined with future data on the galactic cosmic ray actinide composition should help
confirming that galactic cosmic rays are not fresh supernova ejecta. They should also provide a tool
to discriminate between two competing models for the cosmic ray acceleration, one calling for an
isolated supernova exploding in the ordinary old interstellar medium, and one envisioning a
superbubble instead.
\keywords{nuclear reactions, nucleosynthesis, abundances -- solar system: general}}

\maketitle
 
\section{Introduction}

Actinides enter astrophysics in different ways. First, the use of \chem{232}{Th},
\chem{235}{U} and \chem{238}{U} to estimate astrophysical ages has a long history, a
milestone of which is the much celebrated piece of work of Fowler
\& Hoyle (1960). For long, the field of nucleo-cosmochronology that has emerged
from this paper has been aiming at the determination of the age $T_{\rm nuc}$ of
the nuclides from abundances in the material making up the bulk of the solar
system. If indeed the composition of this material witnesses the long history of
the compositional evolution of the Galaxy prior to the isolation of the solar
material, a reliable evaluation of
$T_{\rm nuc}$ (clearly a lower bound to the age of the Universe) requires (i) the
identification of radionuclides with half-lives commensurable with estimated
reasonable galactic ages (i.e. $t_{1/2} \ga 10^9$ y), (ii) the construction of
nucleosynthesis models that are  able to provide the isotopic or elemental yields
for these radionuclides, (iii) high quality data for the meteoritic abundances of
the  relevant nuclides, and, last but not least, (iv) the build-up of models
for the evolution of the abundances of these nuclides in the Galaxy,  primarily in
the solar neighbourhood. All these requirements clearly make the chronometric task
especially demanding. While everybody would agree this far, there are different
ways to look at the question. The use of a so-called `model-independent' approach can
at best set limits on $T_{\rm nuc}$ (Schramm \& Wasserburg 1970). On the other hand,
the `exponential model' introduced by  Fowler \& Hoyle (1960) has been advocated by
many over the years. It takes the stand that, given the presumed complexity of the
chemical evolution of the galactic disk, it is by far preferable to describe its
nucleosynthetic history by a simple function with some adjustable parameters.
In contrast, it is considered by some that it is really worth studying
nucleo-cosmochronology in the framework of models for the chemical evolution of the Galaxy
in the solar neighbourhood which imperatively satisfy as many
observational constraints as possible (e.g. Yokoi et al. 1983; Takahashi 1998).

The astrophysical importance of Th and U has been enhanced further with the 
observation of Th in some very metal-poor stars and of U in one of
them (Sneden et al. 1996; Cayrel et al. 2001). These measurements have raised the hope of 
a possible nuclear-based evaluation of the age of individual stars other than the Sun.

Other recent observational advances have triggered substantial interest in other
actinides which are shorter-lived than Th and U. This comes about following the
measurement with unprecedented resolution of the Galactic Cosmic Ray (GCR) abundances
of the $Z > 70$ elements, including the actinides, using the Trek detector (Westphal
et al. 1998). Further significant progress is expected in the determination of the GCR
abundances of the actinides Th, U, Pu and Cm both with respect to each
other and with respect to the Pt-group of elements. This is hoped to be
achieved with the Extremely Heavy Cosmic Ray Composition Observer (ECCO), a detector
similar to Trek currently under study for deployment on the International Space
Station (Westphal et al. 2000). Precise abundance measurements of this type
would  yield an estimate of the time elapsed between the nucleosynthesis
of the GCR actinides and their acceleration to GCR energies (the
GCR actinide propagation time after acceleration is very short, i.e of the order of 2 My).
Hence, they would help determining whether GCRs were accelerated out of fresh
supernova ejecta, superbubble material, or old well-mixed galactic material.

Let us also mention the attempts to measure the local interstellar medium (ISM)
\chem{244}{Pu} content, which may have some interesting astrophysics implications. At
present, this can be done through the analysis of dust grains of identified interstellar
origin recovered in deep-sea sediments (e.g. Paul et al. 2001). In a near future, the
determination of elemental and isotopic composition of the ISM grains will be a major
goal of research with their recovery to Earth by the Stardust mission (Brownlee et
al. 1996). 

In all the fields referred to above, a necessary condition to interpret the observational
data is to have at disposal r-process predictions for the production ratios
at the sources of the actinides with half-lives typically in excess of about $10^6$ y,
as well as ratios of these actinides to lower $Z$-element abundances. Most importantly, {\it fair
estimates of the uncertainties in these predicted abundances have also to be evaluated}. Providing
such a key information to the cosmo-chemistry, GCRs or interstellar dust 
practitioners is the main aim of this paper (Sect.~2). Attempts to derive constraints on the
predictions of the actinide production from the solar system r-nuclide content and from abundance
measurements in very metal-poor stars are discussed in Sects.~3 and 4. Brief considerations
concerning the relations between our predictions and the GCR actinide content are presented in
Sect.~5. Some conclusions are drawn in Sect.~6. At this point, we want to make clear that a
detailed discussion of the impact of our nucleosynthesis predictions in the fields of
nucleo-cosmochronology, GCR physics and interplanetary actinide content is largely
 out of the scope of this paper, and will possibly be the main concern of other works. 

\section{Actinide production ratios}

\subsection{The multi-event canonical model of the r-process}

The r-process remains the most complicated nucleosynthetic process to
model from the astrophysics as well as nuclear physics points of view (for a review
see Arnould \& Takahashi 1999). On the nuclear physics
side, the nuclear structure properties (such as the nuclear masses, deformation,
\dots) of thousands of nuclei located between the  valley of $\beta$-stability and
the neutron drip line have to be known, as well as their interaction properties, i.e
the ($n,\gamma$) and ($\gamma, n$) rates, $\alpha$- and $\beta$-decay half-lives and
fission probabilities.  Despite much recent experimental effort, those quantities
for almost all the nuclei involved in the r-process remain unknown, so that they have to
be extracted from theory, and are subject to unavoidable uncertainties. 
On top of these nuclear  difficulties, the question of the astrophysical conditions
under which the r-process can develop is far from being settled, all the proposed
scenarios facing serious problems. For this reason, only parametric approaches, such
as the  so-called canonical model (Seeger et al. 1965) can be used to estimate the
actinide  production. The canonical model assumes that some stellar material composed
solely of iron nuclei is subjected to neutron densities and temperatures that remain
constant over the whole neutron irradiation time.  Each event is therefore characterized by
astrophysical conditions that are viewed as free parameters (temperatures $T$, neutron densities
$N_{\rm n}$ and  neutron irradiation times $t_{\rm irr}$, which can be replaced by the related
quantity
$n_{\rm cap}$, the average number of neutrons captured per iron seed). Their values are determined
from a fit to the solar system composition of  the abundances calculated for each canonical event
(CEV).

In this paper, we use the multi-event model introduced by Bouquelle et al. (1996) (see
also Goriely \& Arnould 1997). \footnote{As an unfortunate misconception persists in
part of the r-process literature, we repeat here that the terminology "multi-event" does
not refer necessarily to {\it numerous stars} (like supernovae) responsible for the
production of r-nuclides, but rather to {\it numerous} CEVs characterized by different
thermodynamic conditions. Such a suite of CEVs might for exemple well be associated to
adjacent layers of a {\it single} supernova.} In view of our very poor knowledge of the
precise astrophysical conditions of occurence of a given r-process, the suite of
CEVs needed to approximate the yields of this process is clearly unknown as
well. This is even more true if different types of r-process episodes have to be considered, at
least if the assumption of the `universality' of the r-process yields is not adopted from
the start (see Sect.~4). The multi-event model relies on an iterative inversion
procedure in order to find the ensemble of CEVs which best fits a given
abundance distribution, and in particular the solar system one, for a given nuclear
input. Any change in this nuclear input translates in this approach into a different
ensemble of CEVs in order to fit at best a given observed r-nuclide
distribution. This is thus a unique and efficient tool, in particular to carry out a
systematic study of the impact of nuclear uncertainties on the yield
predictions (Goriely 1999). Some would argue that the multi-event model has a real
drawback because it can mask nuclear structure effects by introducing spurious CEVs. We
would certainly concur with this criticism if indeed one would be able to distinguish at
this point spurious CEVs from real ones. This ambiguity is made even more serious as our
multi-event model as well as other more classical approaches (e.g. Cowan et al. 1999)
make use of the CEV oversimplification. So far, we thus have to live with our inability to
make a clear distinction between astrophysics and nuclear deficiencies of
the r-process model (Goriely \& Arnould 1997).

The calculations referred to as `standard' in the following (see Table~1) are performed with the
adoption of CEVs characterized by the astrophysical conditions (hereafter called SET1) 
 $1.3 \le T_9 \le 1.7$ ($T_9$ is the temperature in $10^9$ K), and $10^{22} \le
N_{\rm n} [{\rm cm}^{-3}] \le10^{29}$.  The CEVs are evaluated for the usual $10 \le n_{\rm cap}
\le 200$ range needed to produce elements between \chem{56}{Fe} and the actinides (only CEVs
with reasonable timescales
$t_{\rm irr}<2$~s are considered). Note that the yields from each of the considered CEVs are not
calculated under the waiting point approximation (in  contrast to Goriely
\& Clerbaux 1999). In these non-equilibrium conditions, the solution of a full reaction
network is made necessary.  When not available experimentally, the nuclear properties
(in particular masses) are taken from the HFBCS-1 model of Goriely et al. (2001), the
neutron capture and photodisintegration rates from Hauser-Feshbach calculations based on
the HFBCS predictions (Goriely 2000), and from the gross theory (GT2) of Tachibana et
al. (1990) for $\beta^-$decays and $\beta$-delayed neutron emissions. In addition, the
spontaneous and $\beta$-delayed fission channels are included at times $ t > t_{\rm irr}$ only.
The various fission probabilities (as well as $\alpha$-decays) are calculated according to the
approximate prescriptions of Kodoma \& Takahashi (1975) with the experimental or ETFSI fission
barriers (Mamdouh et al. 2001). In contrast, neutrino interactions are neglected. It is indeed
meaningless to introduce them in a calculation which does not rely on a detailed astrophysical
site.

When fitting a given observationally-based r-abundance distribution, the multi-event
procedure is the same as the one described by Bouquelle et al. (1996), except that each nuclide
is now given a weight inversely proportional to the uncertainty that is considered to affect
its r-abundance. In the case of the solar system, its r-nuclide content and associated
uncertainties are discussed in detail by Goriely (1999). The data he derives define the solar
system r-abundance set referred to in the following as `SOL1' (see Table~1). The multi-event fit to
SOL1 obtained with the use of the standard calculations referred to above is displayed in Fig.~1.
It involves a superposition of CEVs each of which being responsible for the production of a
limited ensemble of r-nuclides. They can be characterized by the classically used quantity
$S_a^0({\rm MeV}) = \left( 34.075 - \log N_n + {3\over 2} ~\log T_9 \right) T_9 / 5.04$. SET1 
includes CEVs for which $1.4 \la S_a^0({\rm MeV}) \la 4.2$. It is seen that a distribution of 
$S_a^0$ decreasing roughly from 3.5 to 2 MeV with increasing $A$-values is needed in order to
reproduce the solar system distribution of r-nuclides. Case 1 of Table~1 provides the production of
the actinides of astrophysical interest resulting from the fit of Fig.~1. 

\begin{figure} 
\centerline{\epsfig{file=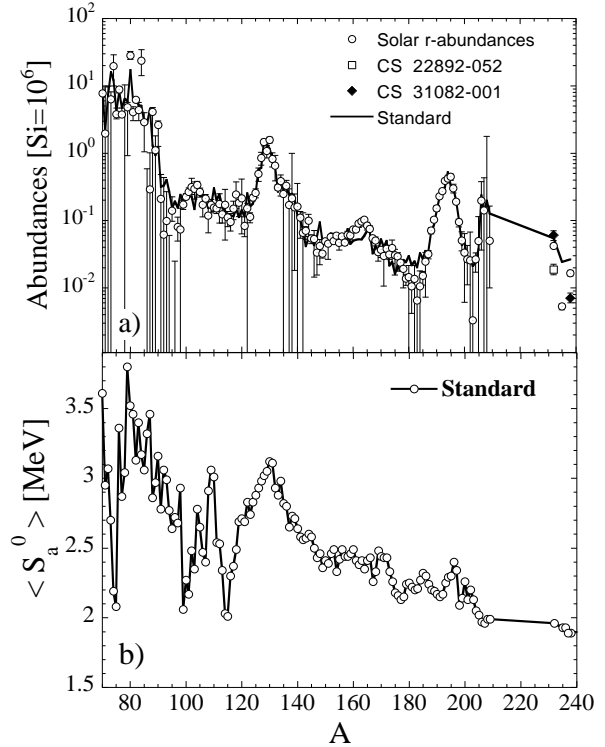,height=10cm,width=8cm}}
\caption{a)  Fit to the set of solar system r-abundances  SOL1 derived from the standard
calculations defined in the main text. The open square corresponds to the Th abundance observed
in CS 22892-052 (Sneden et al. 1996), and the full diamonds to Th and U in CS31082-001 (Cayrel et
al. 2001); b) Average values of $S_a^0$ leading to the fit shown in a).}
\label{fig01}
\end{figure}

\subsection{How uncertain are the actinide abundance predictions?}

The predicted production of the actinides is obviously sensitive to the many, and still more or
less uncertain, astrophysics and nuclear physics aspects of the r-process modelling. For all
fields of space sciences where actinides enter, and in particular for cosmochronological
purposes and for GCR astrophysics, it is of fundamental importance to know to what extent these
uncertainties transpire into the calculated actinide synthesis. In fact, the problem of
their production is particularly acute. There is no stable nuclide heavier than $^{209}{\rm
Bi}$, so that the actinide production can only be constrained by the stable nuclides lying some 30
mass units below. Moreover, in the simple r-process model considered here, the CEVs responsible for
the actinide production do not contribute to the synthesis of the elements lighter than Pb, including
the third Os-Ir-Pt r-peak (Goriely \& Clerbaux 1999). Consequently, the estimated actinide
production can be constrained solely by a fit to the Pb and Bi abundances. 

We evaluate the uncertainties attached to the synthesis of the actinides by focussing on the
solar system r-nuclide content. The conclusions derived from such a study can be generalized to
other cases, and in particular to some very metal-poor stars, at least if the hypothesis of
`universality' of the r-process holds (Sect.~4). In this analysis of the reliability of the
actinide production, the multi-CEV model is adopted throughout. In other words, we put under the
rug all the uncertainties of purely astrophysical nature affecting the r-process.
 This procedure finds its justification in the fact that these astrophysical uncertainties are
almost impossible to quantify today, as the precise site(s) where the r-process develops is(are)
largely unknown.   Sticking to the multi-CEV model, we can, however, provide a limited
evaluation of the astrophysical uncertainties by adopting a second set of conditions for the CEVs:
$T_9 = 1.35$ and $10^{21} \le N_{\rm n} [{\rm cm}^{-3}] \le10^{28}$. This set, referred to as SET2,
is made of a much smaller number of CEVs than SET1, their $S_a^0$ values ranging from about 1.7 to
3.5 MeV. 

A second potential source of uncertainties relates to the fact
that the relative s- and r-contributions to the solar Pb and Bi remain obscure. This increases the
difficulty to extrapolate abundance predictions reliably to the actinides. To estimate the
uncertainty in the solar Pb and Bi r-abundances, we consider not only the set  SOL1 defined above,
but also the set referred to as  SOL2 derived by K\"appeler et al. (1989) from the classical
s-process. The major difference between SOL1 and SOL2 lies in the predicted r-process
contribution to the solar system \chem{208}{Pb} abundance: SOL2 and SOL1 predict values of 0.24 and
in the (0 to 1.78) range, respectively (these r-abundances are expressed in the Si = $10^6$ scale).
The  SOL1 predictions are so uncertain that the \chem{208}{Pb} abundance cannot provide any
meaningful constraint in the CEV fitting procedure.
 
Another important source of uncertainties in the actinide abundance predictions is of nuclear
physics nature. In order to evaluate its extent, we consider modifications to the various
pieces of our standard nuclear input that are known to affect drastically the r-process
predictions (Goriely \& Clerbaux 1999). These concern the nuclear masses, neutron capture and
$\beta$-decay rates, as well as the probabilities of the fission channels that may open at one
point or another during the r-process. With respect to the standard input, we replace 
\begin{itemize}
\item the inclusion of fission just at times $t > t_{\rm irr}$ by the due consideration of the
fission processes during the neutron irradiation as well. This modification is called
`Fission' in Table~\ref{tab1}. In such conditions, the neutron-induced fission channel opens up
on top of those already considered in the standard case;
\item the HFBCS-1 nuclear masses by the following models: (i) HFB-1 (Samyn et al. 2001),
(ii) ETFSI2 (Goriely 2000), (iii) ETFSI-Q (Pearson et al. 1996) which takes into
account in a purely phenomenological way the still-debated strong quenching of the shell
effect found in some microscopic calculations of highly neutron-rich nuclei, (iv) FRDM (M\"oller
et al. 1995), or (v) the Duflo \& Zuker (1995; DZ) prescription based on a very different
approach than the previously cited models, and which has proven its remarkable ability to
predict experimentally known masses;  

\item the GT2 $\beta$-decay rates by (i) the FRDM+QRPA predictions of M\"oller et al.
(1997), or (ii) the  ETFSI+cQRPA estimates of Borzov \& Goriely (2000);

\item the Hauser-Feshbach estimates of the neutron capture rates by the direct  plus
compound model (CN+DC) of Goriely (1998) which takes into account the possible
existence of a low-lying E1 pygmy resonance and the contribution of direct captures of
relevance for exotic neutron-rich nuclei. 
\end{itemize} 

Many studies have compared the quality of these nuclear models on grounds of available
experimental data, as well as the differences in their predictions far away from the valley of
$\beta$-stability. Their impact on the r-process nucleosynthesis has also been analyzed in
various papers (e.g. Goriely 1998, 2001, Borzov \& Goriely 2000). Here, we
restrict ourselves to an analysis of their predictions regarding the actinide synthesis.

Table~\ref{tab1} compares the abundances of Pb and of the actinides with half-lives $t_{1/2} \ga
10^6$ y predicted by the multi-event calculations for the CEV  SET1 and SET2 and for the
solar r-abundances from  SOL1 and  SOL2. These different choices are mingled with the standard
nuclear physics input defined in Sect.~2.1, or with various other combinations of nuclear models
defined above. In all cases, the derived fits to the solar system content of r-nuclides up to Bi
are of more or less equal quality (Goriely \& Clerbaux 1999). The corresponding abundance
\chem{232}{Th_f} of \chem{232}{Th} after decay of its shorter-lived progenitors \chem {236}{U} and
\chem{244}{Pu} (\chem{232}{Th_f} = \chem{232}{Th} + \chem{236}{U} + \chem{244}{Pu}), as well as
the ratios $R_{232,238} =$ \chem{232}{Th_f}/\chem{238}{U} and
$R_{235,238} =$ (\chem{235}{U} + \chem{247}{Cm})/\chem{238}{U} are displayed in Table~\ref{tab2}. 
The considered large variety of nuclear models goes without any doubt along with large ranges of
actinide predicted abundances. 

\begin{table*}
  \caption{Abundances (normalized to Si=$10^6$) of Pb and of the actinides with
half-lives $t_{1/2} > 10^6$ y predicted by multi-event calculations with the astrophysical CEV sets SET1
or SET2  and solar system r-abundances  SOL1 or SOL2. The calculations are also based on the various
indicated combinations of nuclear inputs concerning masses, fission, $\beta$-decay and reaction
rates (see text for more details). For each nuclide, the minimum and maximum abundances are
underlined. The last three lines correspond to the recommended abundances with an estimate of
the minimum and maximum values based on a selection of the above calculations, as described in
 Sect.~4.  The selected cases are marked in bold in the first column.}

   \begin{tabular}{|c|ccc|cccccccc|}
  \hline
Case	&	SET	& SOL &	Nuclear	&	Pb	&	$^{232}$Th	&	$^{235}$U	&	$^{236}$U	&	$^{238}$U	&
$^{237}$Np &
$^{244}$Pu 	&	$^{247}$Cm	\\
\hline																					
 	1	 	&	1	&	1	&	Standard	&	5.66E-01	&	2.47E-02	&	2.19E-02	&	2.27E-02	&	2.67E-02	&	1.84E-02	&	7.28E-03	&	2.47E-03	\\
{\bf 	2	}	&	1	&	1	&	Fission	&	5.75E-01	&	2.53E-02	&	2.26E-02	&	2.35E-02	&	2.44E-02	&	1.90E-02	&	7.43E-03	&	2.57E-03	\\
 	3	 	&	1	&	1	&	HFB-1	&	6.82E-01	&	2.16E-02	&	2.21E-02	&	2.11E-02	&	2.57E-02	&	1.56E-02	&	1.01E-02	&	2.39E-03	\\
{\bf 	4	}	&	1	&	1	&	ETFSI2	&	5.09E-01	&	3.11E-02	&	2.86E-02	&	2.96E-02	&	3.52E-02	&	2.45E-02	&	8.67E-03	&	4.58E-03	\\
{\bf 	5	}	&	1	&	1	&	ETFSI-Q	&	6.07E-01	&	3.05E-02	&	2.82E-02	&	3.31E-02	&	2.46E-02	&	1.49E-02	&	1.89E-02	&	5.84E-03	\\
{\bf 	6	}	&	1	&	1	&	FRDM	&	8.08E-01	&	2.95E-02	&	1.04E-01	&	1.00E-01	&	1.77E-01	&	1.03E-01	&	1.46E-01	&	3.62E-02	\\
 	7	 	&	1	&	1	&	DZ	&	4.65E-01	&	\underline{8.62E-03}	&	8.99E-03	&	9.20E-03	&	1.13E-02	&	7.73E-03	&	5.10E-03	&	1.37E-03	\\
{\bf 	8	}	&	1	&	1	&	QRPA	&	7.74E-01	&	2.77E-02	&	4.15E-02	&	2.94E-02	&	5.66E-02	&	4.63E-02	&	5.01E-02	&	1.09E-02	\\
{\bf 	9	}	&	1	&	1	&	cQRPA	&	7.69E-01	&	2.76E-02	&	4.13E-02	&	2.92E-02	&	5.61E-02	&	4.59E-02	&	4.97E-02	&	1.08E-02	\\
{\bf 	10	}	&	1	&	1	&	CN + DC	&	6.53E-01	&	4.11E-02	&	3.50E-02	&	4.17E-02	&	3.67E-02	&	2.58E-02	&	3.26E-03	&	3.48E-03	\\
{\bf 	11	}	&	1	&	2	&	HFBCS-1	&	7.93E-01	&	4.05E-02	&	5.18E-02	&	5.27E-02	&	9.07E-02	&	6.09E-02	&	2.60E-02	&	1.40E-02	\\
{\bf 	12	}	&	1	&	2	&	HFB-1	&	7.99E-01	&	3.50E-02	&	3.75E-02	&	3.42E-02	&	5.20E-02	&	3.29E-02	&	2.58E-02	&	4.36E-03	\\
{\bf 	13	}	&	1	&	2	&	ETFSI2	&	7.76E-01	&	5.19E-02	&	5.36E-02	&	4.94E-02	&	5.81E-02	&	3.97E-02	&	1.79E-02	&	1.06E-02	\\
 	14	 	&	1	&	2	&	FRDM	&	8.28E-01	&	2.89E-02	&	1.32E-01	&	1.26E-01	&	\underline{2.45E-01}	&	\underline{1.37E-01}	&	\underline{2.28E-01}	&	\underline{5.72E-02}	\\
{\bf 	15	}	&	1	&	2	&	DZ	&	7.53E-01	&	2.75E-02	&	4.51E-02	&	4.48E-02	&	1.00E-01	&	6.01E-02	&	6.52E-02	&	2.20E-02	\\
{\bf 	16	}	&	1	&	2	&	CN + DC	&	8.07E-01	&	6.25E-02	&	5.39E-02	&	6.49E-02	&	6.85E-02	&	4.72E-02	&	6.29E-03	&	6.85E-03	\\
 	17	 	&	2	&	1	&	Standard	&	6.74E-01	&	2.04E-02	&	1.80E-02	&	2.08E-02	&	1.76E-02	&	1.25E-02	&	3.50E-03	&	1.60E-03	\\
 	18	 	&	2	&	1	&	Fission	&	6.60E-01	&	1.96E-02	&	1.71E-02	&	1.97E-02	&	1.51E-02	&	1.18E-02	&	3.32E-03	&	1.49E-03	\\
 	19	 	&	2	&	1	&	HFB-1	&	6.92E-01	&	1.61E-02	&	1.66E-02	&	1.39E-02	&	1.51E-02	&	9.07E-03	&	6.83E-03	&	1.22E-03	\\
{\bf 	20	}	&	2	&	1	&	ETFSI2	&	5.25E-01	&	3.24E-02	&	3.09E-02	&	3.61E-02	&	4.10E-02	&	2.88E-02	&	7.00E-03	&	6.33E-03	\\
{\bf 	21	}	&	2	&	1	&	ETFSI-Q	&	5.49E-01	&	2.66E-02	&	2.44E-02	&	2.99E-02	&	2.71E-02	&	1.46E-02	&	1.84E-02	&	6.78E-03	\\
{\bf 	22	}	&	2	&	1	&	FRDM	&	6.41E-01	&	6.77E-02	&	1.13E-01	&	9.60E-02	&	1.20E-01	&	8.59E-02	&	8.71E-02	&	2.30E-03	\\
 	23	 	&	2	&	1	&	DZ	&	4.94E-01	&	1.27E-02	&	\underline{7.95E-03}	&	\underline{6.86E-03}	&	\underline{9.42E-03}	&	
\underline{5.82E-03}	&	4.22E-03	&	\underline{1.01E-03} \\ 
{\bf 	24	}	&	2	&	1	&	QRPA	&	\underline{8.69E-01}	&	2.61E-02	&	3.47E-02	&	2.87E-02	&	8.05E-02	&	7.07E-02	&	7.51E-02	&	1.55E-02	\\
 	25	 	&	2	&	1	&	cQRPA	&	\underline{4.11E-01}	&	2.42E-02	&	4.91E-02	&	4.68E-02	&	8.92E-02	&	5.09E-02	&	2.02E-02	&	1.64E-02	\\
{\bf 	26	}	&	2	&	1	&	CN + DC	&	5.65E-01	&	2.87E-02	&	2.87E-02	&	3.66E-02	&	2.32E-02	&	1.85E-02	&	\underline{3.19E-03}	&	2.00E-03	\\
{\bf 	27	}	&	2	&	2	&	HFBCS-1	&	7.70E-01	&	2.99E-02	&	3.87E-02	&	4.54E-02	&	5.55E-02	&	3.55E-02	&	1.07E-02	&	9.28E-03	\\
{\bf 	28	}	&	2	&	2	&	HFB-1	&	7.75E-01	&	2.72E-02	&	2.42E-02	&	2.15E-02	&	2.88E-02	&	1.72E-02	&	1.57E-02	&	3.47E-03	\\
{\bf 	29	}	&	2	&	2	&	ETFSI2	&	7.88E-01	&	5.90E-02	&	6.10E-02	&	7.07E-02	&	8.58E-02	&	5.94E-02	&	1.51E-02	&	1.31E-02	\\
 	30	 	&	2	&	2	&	FRDM	&	7.56E-01	&	\underline{9.05E-02}	&	\underline{1.68E-01}	&	\underline{1.41E-01}	&	1.80E-01	&	1.30E-01	&	1.37E-01	&	2.78E-03	\\
{\bf 	31	}	&	2	&	2	&	DZ	&	7.33E-01	&	4.88E-02	&	5.07E-02	&	4.13E-02	&	9.14E-02	&	5.56E-02	&	6.30E-02	&	1.81E-02	\\
{\bf 	32	}	&	2	&	2	&	CN + DC	&	8.14E-01	&	4.83E-02	&	7.24E-02	&	9.79E-02	&	5.54E-02	&	4.64E-02	&	3.95E-03	&	6.09E-03	\\
	\hline																								
	Rec		&	1	&	1	&	Fission	&	5.75E-01	&	2.53E-02	&	2.26E-02	&	2.35E-02	&	2.44E-02	&	1.90E-02	&	7.43E-03	&	2.57E-03	\\
	Min		&		&		&	         	&	5.09E-01	&	2.53E-02	&	2.26E-02	&	2.15E-02	&	2.32E-02	&	1.46E-02	&	3.19E-03	&	2.00E-03	\\
	Max		&		&		&		         &	8.69E-01	&	6.77E-02	&	1.13E-01	&	1.00E-01	&	1.77E-01	&	1.03E-01	&	1.46E-01	&	3.62E-02	\\
\hline																					
\end{tabular}
\label{tab1}
\end{table*}

\section{Can the solar system r-nuclide content constrain the range of predicted actinide
abundances?}

The long-lived \chem{232}{Th}-\chem{238}{U} and \chem{235}{U}-\chem{238}{U}  
pairs have been classically used to estimate the age of the r-nuclides (assumed to be roughly
equal to the age of the Galaxy) from the present meteoritic content of these nuclides.
The opinion has been expressed at several occasions that these pairs have just limited
chronometric virtues (e.g. Yokoi et al. 1983, Arnould \& Goriely 2001). This opinion does
not relate only to the uncertainties in the production ratios exemplified in
Table~\ref{tab2}, which could be increased still further if the oversimplification coming
from the considered CEVs was removed. An additional source of worry comes from the still
large uncertainties affecting the meteoritic Th and U abundances, which amount to at
least 25\% and 8\%, respectively (Grevesse et al. 1996). Last but not least, further
problems arise because of the necessity of introducing the solar system-based
nucleo-cosmochronology in chemical evolution models of the Galaxy. These models have to
satisfy in the best possible way as many astronomical observables as possible. In addition, their
internal consistency has to be checked by comparing the deduced actinide abundance ratios at the
time of formation of the solar system with those adopted at the nucleosynthetic source. In fact,
this consistency requirement is far from being trivial to fulfil. From the construction of a
galactic evolution model generalized in order to include chronometric pairs, Yokoi et al. (1983)
conclude first that the predicted (\chem{235}{U}/\chem{238}{U})$_0$ and
(\chem{232}{Th}/\chem{238}{U})$_0$ ratios at the time $T_\odot$ of isolation of the solar system
material from the galactic one about 4.6 Gy ago is only very weakly dependent on galactic ages, at
least in the explored range from about 11 to 15 Gy. This results largely from the expected rather
weak time dependence of the stellar birthrate (except possibly at early galactic epochs, but a
reliable information on these times is largely erased by the subsequent long period of chemical
evolution). In this situation, the \chem{232}{Th}-\chem{238}{U} and
\chem{235}{U}-\chem{238}{U} pairs are unable to provide chronometric information which cannot be
revealed by other methods. At best, they can provide results in agreement with
conclusions derived from other techniques. As Yokoi et al. (1983) show, this is true at least if the
r-process production ratios lie in the approximate ranges 1 $<$
\chem{235}{U}/\chem{238}{U} $<$ 1.5 and 1.5 $<$ \chem{232}{Th_f}/\chem{238}{U} $<$ 2. If this is
not the case, the adopted galactic evolution model simply does not provide any chronometric
solution in the explored 11 to 15 Gy age range. As seen in Table~\ref{tab2}, these
constraints cannot be satisfied by any of the considered cases. The situation would not be as
desperate if the lower and upper limits of the production ranges $R_{232,238}$	
and	$R_{235,238}$ were stretched by a value of 0.1. Considering the intricacies of the galactic
model adopted by Yokoi et al. (1983) to derive their chronological results, this small extension
certainly does not hurt unsupportably their results. With such a procedure, we would be left with
cases 1,3,4,7,8,9,13,20,22 and 30 of Tables~\ref{tab1} and \ref{tab2}.
 
\begin{table*}
  \caption{Total \chem{232}{Th_f} = \chem{232}{Th} + \chem{236}{U} +\chem{244}{Pu},
$R_{232,238} =$  \chem{232}{Th_f}/\chem{238}{U}, $R_{235,238}$
= (\chem{235}{U}+\chem{247}{Cm})/\chem{238}{U}, and ages
$T^*_{\rm{U,Th}}$ and $T^*_{\rm{U,Eu}}$ (in Gy) of CS 31082-001 based on the U/Th and U/Eu
cosmochronometries. The minimum and maximum values for each entry are underlined. Recommended
ranges of values of these quantities are also provided (see Sect.~4 for details). As in Table~1, the
selected cases are marked in bold in the first column.}

   \begin{tabular}{|c|ccc|ccccc|}
  \hline

	Case		&	set	&	sol	&	Nuclear	&	$^{232}{\rm Th}_f$	&	$R_{232,238}$	&	$R_{235,238}$	&	$T^*_{\rm U,Th}$	&	$T^*_{\rm U,Eu}$	\\
	\hline																		
 	1	 	&	1	&	1	&	Standard	&	0.0547	&	2.05	&	0.91	&	13.55	&	8.38	\\
{\bf 	2	}	&	1	&	1	&	Fission	&	0.0561	&	2.30	&	1.03	&	12.48	&	7.81	\\
 	3	 	&	1	&	1	&	HFB-1	&	0.0528	&	2.05	&	0.95	&	13.54	&	8.14	\\
{\bf 	4	}	&	1	&	1	&	ETFSI2	&	0.0694	&	1.97	&	0.94	&	13.92	&	10.16	\\
{\bf 	5	}	&	1	&	1	&	ETFSI-Q	&	0.0826	&	\underline{ 3.36}	&	1.39	&	\underline{ 8.94}	&	7.86	\\
{\bf 	6	}	&	1	&	1	&	FRDM	&	0.2760	&	1.56	&	0.79	&	16.14	&	20.52	\\
 	7	 	&	1	&	1	&	DZ	&	\underline {0.0229}	&	2.03	&	0.92	&	13.66	&	2.88	\\
{\bf 	8	}	&	1	&	1	&	QRPA	&	0.1072	&	1.89	&	0.93	&	14.30	&	13.20	\\
{\bf 	9	}	&	1	&	1	&	cQRPA	&	0.1064	&	1.90	&	0.93	&	14.30	&	13.15	\\
{\bf 	10	}	&	1	&	1	&	CN + DC	&	0.0860	&	2.34	&	1.05	&	12.31	&	10.43	\\
{\bf 	11	}	&	1	&	2	&	HFBCS-1	&	0.1192	&	1.31	&	0.73	&	17.73	&	16.23	\\
{\bf 	12	}	&	1	&	2	&	HFB-1	&	0.0950	&	1.83	&	0.80	&	14.65	&	12.67	\\
{\bf 	13	}	&	1	&	2	&	ETFSI2	&	0.1192	&	2.05	&	1.10	&	13.56	&	13.38	\\
 	14	 	&	1	&	2	&	FRDM	&	\underline{ 0.3832}	&	1.56	&	0.77	&	16.11	&	\underline{ 22.60}	\\
{\bf 	15	}	&	1	&	2	&	DZ	&	0.1376	&	1.37	&	0.67	&	17.31	&	16.86	\\
{\bf 	16	}	&	1	&	2	&	CN + DC	&	0.1337	&	1.95	&	0.89	&	14.02	&	14.43	\\
 	17	 	&	2	&	1	&	Standard	&	0.0447	&	2.53	&	1.11	&	11.58	&	5.73	\\
 	18	 	&	2	&	1	&	Fission	&	0.0426	&	2.82	&	1.23	&	10.57	&	4.74	\\
 	19	 	&	2	&	1	&	HFB-1	&	0.0368	&	2.43	&	1.18	&	11.97	&	4.75	\\
{\bf 	20	}	&	2	&	1	&	ETFSI2	&	0.0754	&	1.84	&	0.91	&	14.57	&	11.14	\\
{\bf 	21	}	&	2	&	1	&	ETFSI-Q	&	0.0749	&	2.76	&	1.15	&	10.77	&	8.49	\\
{\bf 	22	}	&	2	&	1	&	FRDM	&	0.2508	&	2.09	&	0.96	&	13.39	&	18.03	\\
 	23	 	&	2	&	1	&	DZ	&	0.0238	&	2.53	&	0.95	&	11.59	&	1.71	\\
{\bf 	24	}	&	2	&	1	&	QRPA	&	0.1299	&	1.61	&	\underline{ 0.62}	&	15.81	&	15.46	\\
 	25	 	&	2	&	1	&	cQRPA	&	0.0911	&	\underline{ 1.02}	&	0.73	&	\underline{ 20.09}	&	16.12	\\
{\bf 	26	}	&	2	&	1	&	CN + DC	&	0.0685	&	2.96	&	1.33	&	10.13	&	\underline{ 7.47}	\\
{\bf 	27	}	&	2	&	2	&	HFBCS-1	&	0.0860	&	1.55	&	0.87	&	16.18	&	13.08	\\
{\bf 	28	}	&	2	&	2	&	HFB-1	&	0.0644	&	2.24	&	0.96	&	12.74	&	8.87	\\
{\bf 	29	}	&	2	&	2	&	ETFSI2	&	0.1448	&	1.69	&	0.86	&	15.38	&	15.87	\\
 	30	 	&	2	&	2	&	FRDM	&	0.3681	&	2.04	&	0.95	&	13.61	&	20.64	\\
{\bf 	31	}	&	2	&	2	&	DZ	&	0.1532	&	1.68	&	0.75	&	15.45	&	16.28	\\
{\bf 	32	}	&	2	&	2	&	CN + DC	&	0.1501	&	2.71	&	\underline{ 1.42}	&	10.94	&	13.06	\\
	\hline																		
	Rec		&	1	&	1	&	Fission	&	0.0561	&	2.30	&	1.03	&	12.48	&	7.81	\\
	Min		&		&		&		         &	0.0561	&	1.31	&	0.62	&	8.94	&	7.47	\\
	Max		&		&		&		         &	0.2760	&	3.36	&	1.42	&	17.73	&	20.52	\\
\hline																	
\end{tabular}
\label{tab2}
\end{table*}

\section{Can the observation of very metal-poor stars constrain the range of  
predicted actinide abundances?}

One might also confront the predicted range of actinide productions reported in Tables~\ref{tab1}
and \ref{tab2} with the observations of r-nuclides in old metal-poor stars. Compared with the
case of the solar system discussed in Sect.~3, this chronometry has the advantage of
allowing the economy of a galactic evolution model. Even so, the exercise of finding good reasons
to reject some of the cases considered in Tables~\ref{tab1} and \ref{tab2} is more risky than it
might appear at first to some. The major origin of the difficulties lies in the necessity to make
the assumption that the r-process is `universal'. In other words, the observed patterns of
r-nuclide abundances in metal-poor stars have to be considered as exactly solar.  This is indeed
the only way to take the largest possible advantage of the observed metal-poor star content of Th
and U by bringing them to the status of chronometers.

Before embarking on the problem of deriving constraints, let us first briefly discuss the
validity of the universality hypothesis. In contrast to a widely spread
opinion, Goriely \& Arnould (1997) (see also Goriely \& Clerbaux 1999; Arnould \& Goriely 2001)
consider that {\it the observed convergence of the solar and of the metal-poor stars CS
22892-052 and HD115444 abundance patterns in the $56 \le Z \le 76$ range does in no way
demonstrate the universality of the r-process, without excluding it, however}.   This  conclusion
fully applies as well to the recent abundance determinations of $50 \le Z \le 70$ elements in 22
metal-poor r-process-rich stars (Johnson \& Bolte 2001). In addition, there has been some bad
news for the many proponents of the r-process universality with the observations by Cayrel et al.
(2001) of the very metal-poor r-process-enriched halo star CS 31082-001. While a
large similarity between the $56 < Z < 70$ element patterns in CS 22892-052, HD115444 and CS
31082-001 is reported, the abundances in the latter star differ significantly from those of the
other two stars in the $Z > 70$ range, including Th. In these conditions, the
universality assumption would lead to quite odd chronometric conclusions (Arnould \&
Goriely 2001). In particular, the Th/Eu ratio in CS 31082-001 is about 3.2 times larger
than in CS 22892-052. Hence, under the universality assumption, CS 22892-052 (with
[Fe/H]=-3.1) predates CS 31082-001 (with [Fe/H]=-2.9) by 24 Gy, and would thus be
about 36 Gy old.

The Pb/Th ratios observed in CS 22892-052  (log$\epsilon$(Pb/Th)=1.80~$\pm$~0.40) and in
CS 31082-001 (log$\epsilon$(Pb/Th)$<$0.76) may run down as well the universality
hypothesis. A correlation indeed exists between the r-process production of Pb and Th,
the abundances of these two elements increasing or decreasing concomitantly (see Fig.~1 of
Goriely \& Clerbaux 1999). In these conditions, and if the
universality of the Pb/Th ratio is assumed, the observed Pb/Th values turn out to be discrepant
by a factor of about 10, at least if the two stars have roughly the same age. If this is indeed
the case (which is not a farfetched assumption in view of their similar [Fe/H] ratio),
either the universality assumption is invalid, and a specific actinide-producing
r-process has to be called for, or the Pb in CS 22892-052 is largely of s-process
origin. This could well be the case if a Pb pollution by a low-metallicity
AGB star could be invoked (in particular if CS 22892-052 would be a binary star). As shown by
Goriely \& Siess (2001) and Van Eck et al. (2001), the s-process in  extremely low-metallicity
AGB stars indeed leads almost exclusively to the production of Pb, in such a way that none
of the elements lighter than Pb  would see its abundance affected by the s-process. It
would be of substantial interest to find ways to discriminate between the r-process
non-universality and the s-process pollution scenarios. Even if the assumption of a
universal r-process appears to be more and more fragile with time, we dare {\it suppose} in
the following that it indeed holds in order to examine if constraints can be put in such
a favourable situation on the nuclear and astrophysical models for use in r-process calculations, and
consequently on the actinide production.

To make things clear, the universality assumption means nothing more and nothing
less than the following: {\it all possible combinations of CEVs found in nature always result in
the same final abundance pattern}. It is important to acknowledge that the precise
characteristics of each of the CEVs which may contribute to the universal mix are unknown, as
well as the relative level of the contribution of each of the CEVs to the mix. In such
conditions, it cannot be excluded that different combinations of different individual CEVs lead
to the same final mix. An improper evaluation of this situation may lead to spurious constraints
on the nuclear models to be used as input to the CEV calculations. Another source of spurious
constraints might arise by focussing on the quality of a fit to a single nuclide, even in a quite
crucial region, like the Pb one (e.g Cowan et al. 1999). Such a highly punctual quality of
fit may be quite misleading in the evaluation of the merits of a global nuclear input and of the
associated predictions of the actinide production. 

One might imagine gaining some constraints from Th alone. The ages of CS 22892-052 and  CS
31082-001 derived from a confrontation between the observed Th abundances and those displayed in
Table~\ref{tab1} vary in the $3 \la T^* {\rm[Gy]} \la 60$ and $0 \la T^* {\rm[Gy]} \la 37$ 
ranges, respectively. Obviously, some of these ages are meaningless when reference is made to
other age determination techniques (reviewed in e.g. von Hippel et al. 2001). On such grounds,
one might thus be tempted to eliminate right away some of the nuclear models used in
Table~\ref{tab1}. This stand is perfectly legitimate in the simplistic astrophysics scenario of
superposed CEVs we have adopted (Sect.~2.1). In contrast, by so doing, one may face the danger of
reaching wrong conclusions if more realistic r-process models were considered. The absolute Th
production is indeed likely to be highly dependent on these models. The consequences of this
situation are aggravated by the fact that the ages derived from Th are especially sensitive to
its precise yield predictions, as a result of its long half-life ($t_{1/2} \approx 14$ Gy). 

Let us thus turn to safer ways to derive meaningful constraints. One
can in particular rely on accurate measurements of U/Th in individual low-metallicity stars (Arnould
\& Takahashi 1999, Goriely \& Clerbaux 1999), these two actinides being produced
concomitantly. A U/Th ratio has been reported for the star CS 31082-001 (Cayrel et al. 2001; Hill et
al. 2001).
This is real good news, even if the situation is not free of observational and theoretical
difficulties. The former ones are discussed by Cayrel et al. (2001),  and are vividly
illustrated by the very recent re-evaluation of the ThII and
UII line strengths by Nilsson et al. (2001, 2001a) leading to the revised values of
log~$\varepsilon$(U/H)=-1.83 and log~$\varepsilon$(Th/H)=-0.89. The resulting log(U/Th) for CS
31082-001 is -0.94~$\pm$~0.09 (Cayrel,  private communication) instead of the originally derived
value of -0.74 (Cayrel et al. 2001; Hill et al. 2001). On the theoretical side, uncertainties, if
they cannot of course be eliminated, are largely reduced, however. The predicted stellar ages lie in
the restricted
$9 \la T^* {\rm[Gy]}
\la 18$ range (Table~\ref{tab2}). None of the r-process calculations shown in Table~\ref{tab2} can
be excluded for sure on the basis of the U/Th chronometry.
This situation just translates the enhanced reliability of the predictions of the
U/Th ratio compared to the ones based on a single actinide.

Some constraints on the actinide
production could be gained from the development of chronometries based on other r-process pairs,
ideally to be used in conjunction with U/Th. One of these pairs which has already been adopted
quite often in the past is Th/Eu. This choice has of course been dictated by available
observations, but is in fact quite unfortunate from a theoretical point of view. It indeed
combines the drawback coming from the long Th lifetime, as stressed above, with the inconvenience
that Eu and Th are not produced in the same CEV, so that the Th/Eu production ratio is expected
to be quite drastically model dependent. Considering now the U/Eu pair, it is, of course, 
expected to be as model dependent as Th/Eu.  It has however the pleasing feature of involving U,
which has a lifetime about three times shorter than Th.  This has the
advantage of making chronometric conclusions less sensitive to the precise
production of U than to the one of Th.  So, although it also requires the
assumption of r-process universality, the U/Eu chronometry may be safer than
the Th/Eu one. Table~\ref{tab2}
indicates that the use of U/Eu leads to a  CS 31082-001 age in the $7 \la T^* {\rm[Gy]}
\la 23$ range. Under the constraint that the age of CS 31082-001 obtained from U/Th
($T^*_{\rm{U,Th}}$) and U/Eu ($T^*_{\rm{U,Eu}}$) should be the same,  we reject the
cases for which $T^*_{\rm{U,Th}}$ and $T^*_{\rm{U,Eu}}$ differ by more than 5~Gy. From the retained
cases, we suggest in Tables~1 and 2 recommended, minimum and maximum values for each predicted
abundance and age. Note that U/Os would be a valuable chronometric pair as well, in principle at
least. We do not use it, however. The star CS 31082-001 indeed exhibits a Os overabundance of about
0.35~dex with respect to the universal pattern (Hill et al. 2001).
 This clearly contradicts the universality assumption which is the basis of all the chronometric
considerations making use of metal-poor stars. The measured Os overabundance would imply a  
CS~31082-001 age $T^*_{\rm{U,Os}}$ exceeding $T^*_{\rm{U,Eu}}$ by about 5~Gy. 
More generally, this situation creates some discomfort about the whole cosmochronometric virtues of
the  actinides in metal-poor stars.
  
Some closing remarks are in order at this point. First, we want to restate that the reported
constraints strictly relate to the simplistic CEV model of the r-process, and put totally under
the rug uncertainties related to the very nature of the r-process and the likely intricate
astrophysical conditions that prevail during its development. A detailed treatment of
uncertainties of such a nature is largely out of reach in view of the poor characterization of
the true astrophysical context in which an analysis of this type should have to be conducted.
Second, the constraints adopted to select the recommended actinide productions and their ranges
of variations given in Tables~\ref{tab1} and \ref{tab2}, while admitedly highly subjective, appear
reasonable to the authors only {\it under the assumption of the universality of the r-process}. At
discussed above, this basic assumption appears to be more and more questionable as data accumulate.
As a direct consequence, the derived constraints are increasingly unsecure.  
 
\section{The Galactic cosmic ray actinide composition} 

As stressed in Sect.~1, measurements of actinide abundances relative to each other and to
the Pt-group of elements is of very special interest in order to help determining whether 
GCRs are fresh supernova ejecta, or superbubble material, or old galactic material.

It is generally taken for granted today that supernova explosions are the most probable GCR
energy source. It is believed that individual supernova remnants may be responsible for
the acceleration of external, swept-up ISM, with at most
a very tiny contribution of internal, nucleosynthetically processed material
(Meyer \& Ellison 1999; Ellison \& Meyer 1999). Observation of GCR actinides  could confirm 
this scenario. Massive
star supernova explosions  are not random in the Galaxy, however, and concentrate strongly
in OB associations. In fact, fireworks of sequential explosions of tens of massive stars lasting
for periods as short as a few million years could create `multiple supernova remnants'. These can
grow into so-called `superbubbles' made of hot tenuous plasma most commonly observed from their
x-ray emission in our and nearby galaxies (e.g. Spitzer 1990). Superbubbles might well be
privileged galactic locations for the acceleration of matter to GCR energies (Higdon et al.
1998; Parizot 2000, 2001; Bykov 2001). Just as in the case of isolated remnants, each superbubble 
remnant
accelerates external, swept-up superbubble material. This material,
while predominantly ordinary ISM evaporated from nearby
clouds, is, however, significantly contaminated by the recent ejecta of previous local
supernovae.  In addition, turbulent acceleration should take place steadily
throughout the superbubble gas.  So, more fresh supernova ejecta may be
expected in GCRs from superbubbles than from isolated supernovae.
 This results from
the study of both the superbubble dynamics and from considerations about the synthesis of the
light elements Li, Be and B in the early galaxy (e.g. Parizot 2000, 2001). This increased fraction 
of fresh ejecta also nicely accounts for the GCR \chem{22}{Ne} anomaly (Meynet et al. 2001).

\begin{figure} 
\centerline{\epsfig{file=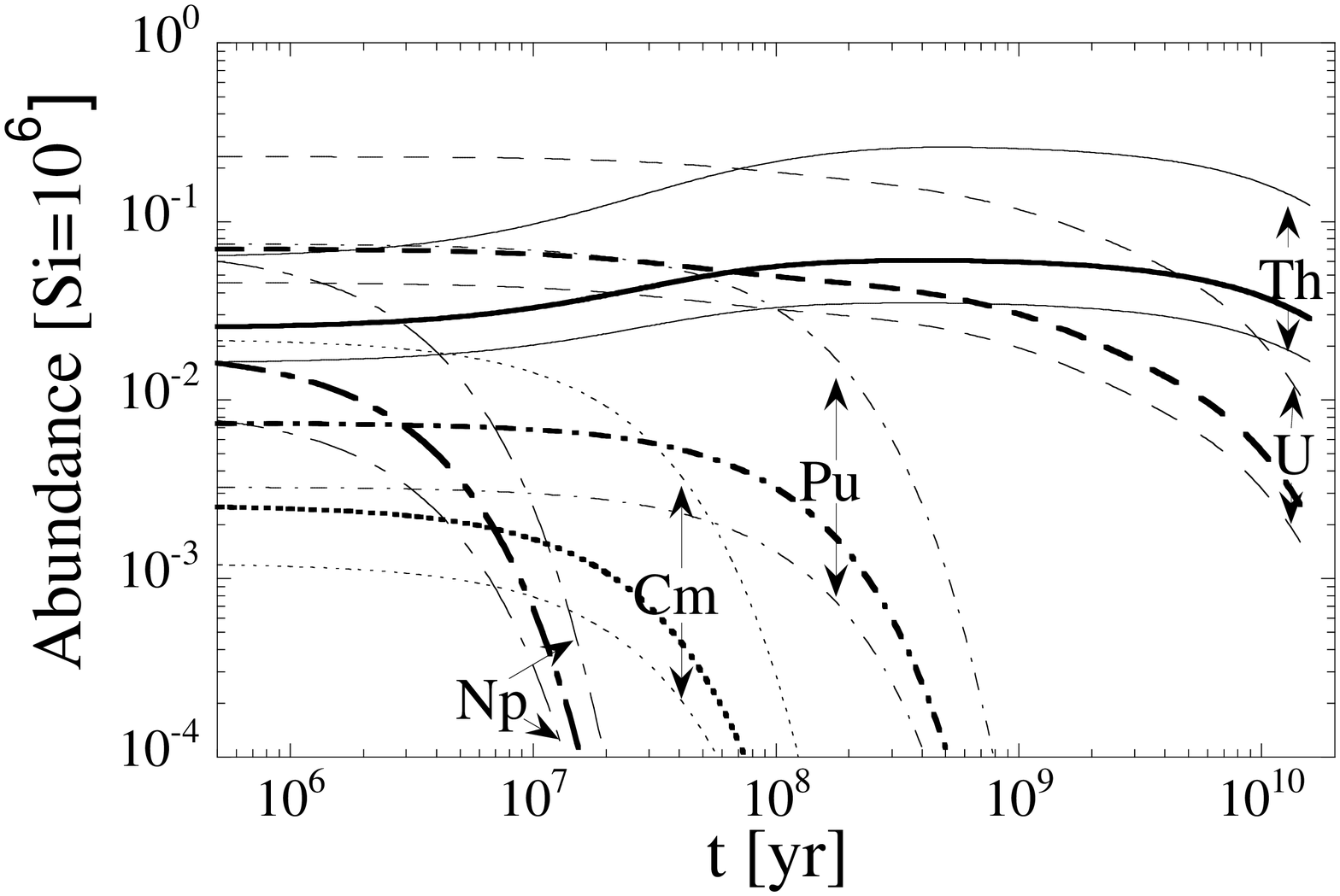,height=6.5cm,width=8cm}}
\caption{Time variations of the elemental abundances of the actinides of interest for GCR
studies. A single r-process production is assumed at time zero. The thick and thin lines
correspond to the recommended values and their lower or upper limits given in
Table~\ref{tab1}.}  
\label{fig02}
\end{figure}
 
An accurate measurement of the actinide GCR composition which is within the reach of the
planned ECCO detector (Sect.~1) could provide an additional and specific way of
distinguishing between the isolated supernova remnant and supperbubble scenarios for the
source of the GCRs (Westphal et al. 2000). Indeed, the relatively short lifetimes of the
superbubbles imply a young age (say $< 30$ My) for the emerging GCRs. Such an age
constraint is relaxed in the case of isolated supernova remnants. So, as noted by Westphal
et al. (2000), GCRs originating from supperbubbles would likely be young enough for
containing a significant amount of \chem{244}{Pu} and \chem{247}{Cm} the lifetimes of which are
commensurable with the superbubble ones. Concomitantly, the presence of \chem{247}{Cm} in the GCRs
and their implied young age would be the indication that Th, U, and Pu have abundance ratios close
to their r-process yields. In such conditions, first hand constraints on the actinide production
by the r-process could be gained, in addition to quality information on the origin and age of the
GCRs.
 
We provide in Fig.~\ref{fig02} some predicted abundances of interest for the quantitative 
interpretation of the future GCR actinide measurements. We retain only the cases of Table~2
satisfying the constraints set in Sect.~4 from the spectroscopic analyses of very metal-poor
stars. A detailed discussion of these results is out of the scope of this paper.
 
\section{Conclusion}

The reliable evaluation of the r-process production of the actinides and careful estimates of
the uncertainties affecting these predictions are key ingredients especially in
nucleo-cosmochronology studies based on the analysis of very metal-poor stars or on the
composition of meteorites. This type of information is also required in order to prepare
making the best possible use of future high precision data on the GCR actinide composition, to
establish a more quantitative confrontation with an expected growing body of measurements of the
actinide content of the interstellar medium in the solar neighbourhood, or even of meteoritic
grains of suspected circumstellar origin. 

This paper provides the practitioners in these various fields with the most detailed and
careful analysis of the r-process actinide production available to-date. These predictions are
based on a version of the multi-event canonical model of the r-process which discards the largely
used waiting point approximation. This approach is especially well suited for
evaluating uncertainties which emerge from the adopted nuclear physics. It is acknowledged that
this model is astrophysically oversimplified. However, going far beyond this
simplification is close to impossible in view of the present lack of knowledge about the site(s)
of the r-process and of the physical conditions prevailing during its development.
Thirty-two different combinations of models for the calculation of nuclear masses,
$\beta$-decay rates, $\beta$-delayed probabilities, as well as fission processes are considered.
Two variants of the model used to predict nuclear reaction rates are adopted. The impact of
uncertainties in the r-process production of the closest Pb and Bi r-nuclides is also analyzed.

For the defenders of the universality of the r-process, we show how this property can help
defining constraints on the actinide yield predictions and associated uncertainties. To
derive these constraints, we take advantage of the simultaneous Th and U measurements in the star CS
31082-001 and we impose somewhat arbitrarily that the age derived from the observed U/Th ratio
should not differ from the U/Eu age by more than 5 Gy.  We acknowledge that other criteria could be
defined on grounds of the metal-poor star actinide data and of the r-process universality. From the
information provided by Tables~1 and 2, the readers may in fact set up their own choice of criteria
and define their corresponding own constraints and recommended actinide productions, if they find
them more appropriate. We stress most strongly that {\it these constraints are just meaningless if
the universality of the r-process does not hold, a situation that we clearly favour}. In fact, the
universality of the r-process appears to be less and less likely as data accumulate.   

We also discuss briefly
constraints that could be put on the r-process nuclear physics input from nucleo-cosmochronological
results based on the confrontation between meteoritic actinide abundances and predictions relying on
a model for the chemical evolution of the Galaxy in the solar neighbourhood. The problem of setting
up this second set of constraints does not relate to the likely non-universality of the r-process, 
but instead to the intricacies introduced by the necessity to rely on models for the chemical
evolution of the Galaxy. As a consequence, providing solid values for the actinide production, or
even reliable `error bars' on this production, remains a highly risky exercise. Clearly, much is
left for the future.
 
Finally, our calculations of the actinide production are shown to provide the necessary tools
for using at best future data about the galactic cosmic ray actinide composition. The
confrontation between these observations and our predictions should help confirming that galactic
cosmic rays are not fresh supernova ejecta. They should also provide a way of discriminating
between two competing models for their acceleration: the isolated supernova
remnant exploding in ordinary, old ISM and the superbubble scenario.
 
\begin{acknowledgements}
We thank R.~Cayrel for bringing to our attention the sill unpublished re-evaluation of the Th and U
abundances in the star CS 31082-001. We also thank J.-P. Meyer and A.J. Westphal for fruitful
discussions concerning the actinide content of GCRs, and for their guidance in the field. J.-P.
Meyer and R.~Cayrel are also acknowledged for their careful reading of the manuscript and for
suggestions which have helped improving the manuscript. S.G. is FNRS Research Associate.
\end{acknowledgements}

\end{document}